\documentclass[10pt]{article}
\usepackage[OE]{express}
\renewcommand{\vec}[1]{\mathbf{#1}} 
\usepackage[]{graphicx}
\usepackage{graphics}
\usepackage{amsmath}
\usepackage{amsfonts}
\usepackage{amssymb}
\usepackage{wasysym} 
\usepackage{nicefrac} 
\usepackage{setspace}
\usepackage{color} 
\usepackage{nicefrac} 

\renewcommand{\Im}{\operatorname{Im}}
\newcommand{\figref}[1]{Fig.~\ref{fig:#1}}

\newcommand{\Figref}[1]{Figure~\ref{fig:#1}}
\renewcommand{\eqref}[1]{Eq.~(\ref{eq:#1})}

\newcommand{\secref}[1]{Sec.~\ref{sec:#1}}
\newcommand{\Secref}[1]{Section~\ref{sec:#1}} 
\newcommand{\Pex}{P_{\text{ex}}}
\newcommand{\Pup}{P_{\text{up}}}

\begin{document}
\title{Near-field thermal upconversion and energy transfer through a Kerr medium: Theory}

\author{Chinmay Khandekar,\authormark{1} Alejandro W. Rodriguez,\authormark{2}}

\address{\authormark{1,2}Princeton University, Princeton, NJ 08544,
  USA}

\email{\authormark{1}cck@princeton.edu} %% email address is required
\email{\authormark{2}arod@princeton.edu}

%%%%%%%%%%%%%%%%%%% abstract and OCIS codes %%%%%%%%%%%%%%%%
%% [use \begin{abstract*}...\end{abstract*} if exempt from copyright]
\begin{abstract}
  We present an approach for achieving large Kerr
  $\chi^{(3)}$--mediated thermal energy transfer at the nanoscale that
  exploits a general coupled-mode description of triply resonant,
  four-wave mixing processes. We analyze the efficiency of thermal
  upconversion and energy transfer from mid- to near-infrared
  wavelengths in planar geometries involving two slabs supporting
  far-apart surface plasmon polaritons and separated by a nonlinear
  $\chi^{(3)}$ medium that is irradiated by externally incident
  light. We study multiple geometric and material configurations and
  different classes of interveening mediums---either bulk or
  nanostructured lattices of nanoparticles embedded in nonlinear
  materials---designed to resonantly enhance the interaction of the
  incident light with thermal slab resonances. We find that even when
  the entire system is in thermodynamic equilibrium (at room
  temperature) and under typical drive intensities $\sim
  \mathrm{W}/\mu\mathrm{m}^2$, the resulting upconversion rates can
  approach and even exceed thermal flux rates achieved in typical
  symmetric and non-equilibrium configurations of vacuum-separated
  slabs. The proposed nonlinear scheme could potentially be exploited
  to achieve thermal cooling and refrigeration at the nanoscale, and
  to actively control heat transfer between materials with
  dramatically different resonant responses.
\end{abstract}

% REPLACE WITH CORRECT OCIS CODES FOR YOUR ARTICLE, MINIMUM OF TWO; Avoid using the OCIS codes for “General” or “General science” whenever possible.
%For a complete list of OCIS codes, visit: https://www.osapublishing.org/oe/submit/ocis/

%%%%%%%%%%%%%%%%%%%%%%% References %%%%%%%%%%%%%%%%%%%%%%%%%
\bibliographystyle{unsrt}

%%%%%%%%%%%%%%%%%%%%%%%%%%  body  %%%%%%%%%%%%%%%%%%%%%%%%%%
\section{Introduction}

The field of nonlinear optics has experienced unprecedented growth in
the last several decades, leading to advances in a wide range of
optical technologies with applications for signal
processing~\cite{koos2009all,cotter1999nonlinear},
detectors~\cite{zernike2006applied,hadfield2009single},
spectroscopy~\cite{mukamel1999principles}, among others. Due to the
inherently weak nature of bulk optical
nonlinearities~\cite{boyd2003nonlinear}, most nonlinear devices rely
on resonant systems e.g. large-etalon
mirrors~\cite{saleh1991fundamentals}, photonic-crystal
defects~\cite{soljavcic2004enhancement} and plasmonic
resonators~\cite{kauranen2012nonlinear}, that confine light into small
mode volumes and over long timescales~\cite{joannopoulos2011photonic},
thereby reducing power requirements. As these power requirements are
scaled down, even relatively small effects stemming from thermal
fluctuations can be altered by material
nonlinearities~\cite{dykman1975theory,kheirandish2011finite}, but such
phenomena have only just begun to be
explored~\cite{khandekar2015radiative,soo2016fluctuational}. For
instance, we recently showed in Ref.~\cite{khandekar2015radiative}
that at high temperatures, optical nonlinearities can alter the
emission spectrum of photonic resonators, leading to assymetric
lineshapes and greater-than-blackbody emission under passive (purely
thermodynamic) operating conditions. In subsequent
work~\cite{khandekar2015thermal}, we showed that an optically driven
photonic resonator can exhibit nonlinear thermal effects at a lower
"temperature scale" as well as lead to new phenomena, including the
appearance of Stokes/anti-Stokes emission lines and thermally
activated transitions~\cite{dykman1975theory}. In this paper, we
extend our earlier work to show that the Kerr $\chi^{(3)}$ nonlinear
response of a passive medium can be exploited to efficiently upconvert
thermal radiation from mid-infrared to near-infrared or visible
wavelengths.

Thermal radiation from hot bodies has been studied for over a century
now~\cite{rytov1988principles,polder1971theory,eckhardt1984macroscopic}. The
recent development and application of powerful numerical techniques
capable of describing thermal radiation from complex, structured
materials has paved the way for the design of emitters with unique and
spectrally selective
properties~\cite{howell2010thermal,granqvist1985spectrally,
  greffet2007coherent}. A signficant body of work has focused on the
study of radiation in the near field (short gap sizes $\ll$ thermal
wavelength $\lambda_T \approx 10\mu$m near room temperature), where
bound surface resonances can contribute heat and cause two objects
held at different temperatures and separated by sub-micron gaps to
exchange heat at rates that are orders of magnitude larger than those
predicted by the Stefan--Boltzmann law (applicable only in the far
field)~\cite{joulain2005surface,basu2009review,
  otey2014fluctuational,jones2013thermal}. This in turn has created
opportunities for potential advances in the areas of
photovoltaics~\cite{de2011research}, imaging~\cite{de2006thermal}, and
more generally, thermal devices operating at the nano
scale~\cite{ben2014near,messina2012graphene,song2015near}. More
recently, there has been interest in achieving active control of
near-field heat transfer~\cite{otey2010thermal}, such as through gain
media~\cite{khandekar2016giant,ding2016active} or via chemical
potentials~\cite{chen2015heat}, with potential applications to
nanoscale thermal
refrigeration~\cite{guha2012near,chowdhury2009chip,shakouri2006nanoscale}. In
this paper, we propose a different active mechanism for tailoring
radiative heat transfer that exploits parametric optical
nonlinearities mediated by externally incident light to extract and
upconvert "thermal energy" trapped in the near field of a planar body
unto another. As a proof of concept, we analyze heat exchange in
planar geometries supporting surface plasmon polaritonic (SPP)
resonances at far-apart wavelengths. In particular, we study planar
configurations of silicon carbide (SiC) slabs separated from either
silver (Ag), potassium (K), or indium tin oxide (ITO) slabs by a
nonlinear (bulk or nanostructured) chalcogenide (ChG) film. Although
the large SPP mismatch of the slabs leads to negligible heat exchange
in the absence of a pump---SiC supports mid-infrared SPPs while the
absorber slabs can support either near-infrared or visible SPPs---we
show that under external illumination, energy transfer across the gap
can be significant. In particular, we consider multiple
interveening-gap designs---either bulk nonlinear thin films or
nanostructured media consisting of lattices of nanoparticles embedded
in a nonlinear medium---and find that even when the entire system is
in thermodynamic equilibrium (at room temperature), the energy flux
rate of mid-infrared photons which get upconverted and subsequently
absorbed at near-infrared or visible wavelengths can be as large as
$10^4~\mathrm{W}/\mathrm{m}^2$ for relatively low pump intensities
$\sim \mathrm{W}/\mu\mathrm{m}^2$, approaching and even exceeding
typical flux rates observed in more commonly studied, non-equilibrium
(passive) scenarios in which the slabs are held at a large temperature
differential. Hence, this scheme allows significant thermal extraction
and absorption of thermal radiation at short wavelengths (otherwise
inaccessible under purely passive scenarios) and between very
different (such as non-resonant) materials. Our calculations suggest
that with additional design optimizations, it may be possible to
exploit this approach for radiative cooling and refrigeration.

The paper is organized as follows. In \secref{CMT}, we consider a
representative and generic system consisting of three resonators
supporting modes at far-apart frequencies $\omega_1, \omega_2$ and
$\omega_3=\omega_1+2\omega_2$. The resonator modes are coupled by a
Kerr $\chi^{(3)}$ medium which is driven at $\omega_2$ by externally
incident light, leading to upconversion of thermal radiation from
$\omega_1$ to $\omega_3$. Our analysis is based on a coupled mode
theory framework~\cite{haus1984waves} that describes the underlying
resonant four-wave mixing process and lays out general geometric and
operating conditions required to maximize upconversion. In
\secref{RHT}, we consider concrete, physical examples of this scheme
based on planar geometries and explore different possible material
choices of emitter and absorber slabs separated by a nonlinear
medium. \Secref{nanospheres} and \secref{nanodisks} describe
situations where the upconversion is mediated by an intermediate
medium consisting of a lattice of nanoparticles (composite core/shell
nanospheres or graphene nanodisks) embedded in ChG, whereas
\secref{bulk} focuses on upconversion through a bulk ChG
film. Finally, in \secref{conclusion} we provide a summary of our main
results and possible directions for future work.
 
\begin{figure}[t!]
  \centering \includegraphics[width=1\linewidth]{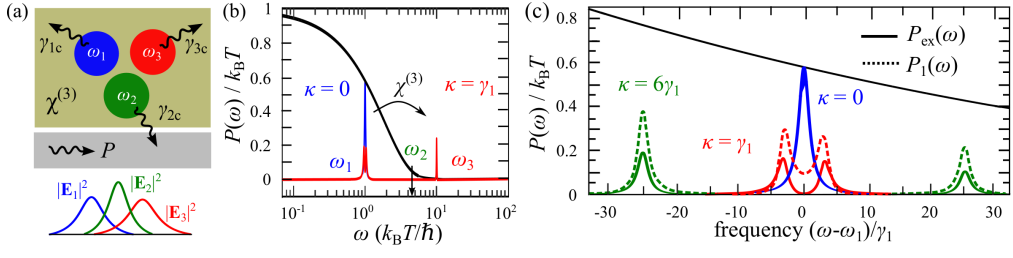}
  \caption{(a) Schematic of three equal-temperature $T$ resonators
    supporting modes at $\omega_{j}$, each with decay rate $\gamma_j$
    stemming from internal dissipation $\gamma_{jd}$ and/or radiation
    into an external channel $\gamma_{jc}$, with $j=\{1,2,3\}$.  The
    thermal modes 1 and 3 are strongly coupled to one another through
    a four-wave mixing process involving a Kerr $\chi^{(3)}$ medium
    excited by externally incident light of power $P$ from a waveguide
    that resonantly couples to the mode at
    $\omega_2=\nicefrac{1}{2}(\omega_3-\omega_1)$.  Nonlinear mixing
    between modes 1 and 3 can be described by an effective, linear but
    power-dependent coupling $\kappa$ [\eqref{nlcoupling}] obtained
    via a spatial overlap between the three modes, given in
    \eqref{beta}. (b) Thermal emission spectra $P_j(\omega)$ (blue and
    red curves) normalized by $k_\mathrm{B} T$, for a choice of
    far-apart frequencies $\omega_1=k_\mathrm{B} T/\hbar$ and
    $\omega_3=10\omega_1$, and decay rates
    $\gamma_{jd}=\gamma_{jc}=0.01\omega_1$, as a function of the
    dimensionless frequency $\hbar \omega / k_\mathrm{B} T$. Emission
    at any $\omega$ is bounded above by the Planck distribution
    $\Theta(\omega,T)$ (black curve) in the absence of the pump
    $\kappa=0$ (blue curves), but is exponentially enhanced under
    finite $\kappa > 0$ (red curves). (c) Thermal emission
    $P_1(\omega)$ (solid lines) and heat-transfer $\Pex(\omega)$
    (dotted lines) spectra near $\omega_1$ as a function of the
    dimensionless frequency $(\omega-\omega_1)/\gamma_1$,
    demonstrating splitting of the resonances into Stokes
    (red-shifted) and anti-Stokes (blue-shifted) peaks, which grow
    apart with increasing $\kappa$ and further enhance emission.}
  \label{fig:cmt}
\end{figure}

\section{Coupled-mode theory}
\label{sec:CMT}

We first illustrate the basic thermal upconversion mechanism by
considering a representative system, depicted in the top inset of
\figref{cmt}(a), involving resonators that support modes at
$\omega_1$, $\omega_2$ and $\omega_3=\omega_1+2\omega_2$, the second
of which is coupled to an external channel. The modes are assumed to
have widely different frequency and therefore do not couple linearly
to one another. They can however interact nonlinearly through a
four-wave mixing process~\cite{boyd2003nonlinear} mediated by a
$\chi^{(3)}$ medium via their field $E_j$ ($j=1,2,3$) overlaps and
initiated by externally incident light at $\omega_2$ from the
channel. Such a system is well described by the following temporal
coupled-mode equations in terms of a few key geometric
parameters~\cite{haus1984waves}:
\begin{align}
  \frac{da_1}{dt} &= (i\omega_1-\gamma_1)a_1 - i\beta\omega_1 a_3
  (a_2^*)^2 + \sqrt{2\gamma_{1d}}\xi_1 \label{eq:a1}
  \\ \frac{da_2}{dt} &= (i\omega_2-\gamma_2)a_2 -i\beta\omega_2 a_3
  a_1^* a_2^* + \sqrt{2\gamma_{2d}}\xi_2 +
  \sqrt{2\gamma_{2c}}s_{\text{in}} \label{eq:a2} \\ \frac{da_3}{dt} &=
  (i\omega_3-\gamma_3)a_3 - i\beta^* \omega_3 a_1 a_2^2 +
  \sqrt{2\gamma_{3d}}\xi_3 \label{eq:a3}
\end{align}
where $a_j$ denotes the mode amplitude of mode $j \in [1,3]$,
normalized so that $|a_j|^2$ is the mode energy, and $\gamma_j =
\gamma_{jd}+\gamma_{jc}$ denotes its decay rate, resulting from either
material dissipation $\gamma_{jd}$ or coupling to external channels
$\gamma_{jc}$ (e.g. radiation or a waveguide). Each mode is assumed to
be in local thermodynamic equilibrium~\cite{khandekar2015radiative}
and hence subject to thermal sources $\xi_j$ satisfying $\langle
\xi_j^*(\omega)\xi_j(\omega') \rangle =\Theta(\omega,T_j)
\delta(\omega-\omega')$, where $\Theta(\omega,T_j) = \hbar
\omega/[\exp(\hbar \omega/k_\mathrm{B} T_j) - 1]$ is the Planck
distribution associated with the local resonator temperature $T_j$ and
$\langle \cdots \rangle$ denotes a thermodynamic, ensemble
average.~\cite{karatzas2012brownian}. A monochromatic coherent drive
incident from the channel, $s_{\text{in}}=s_0 \text{exp}(i\omega_2 t)$
normalized such that $P=|s_0|^2$ denotes the power, facilitates the
nonlinear interaction captured by the following nonlinear coupling
coefficient:
\begin{align}
\beta=\frac{\int dV \chi^{(3)}_{ijkl}
  E_{1i}E_{2j}E_{2k}E^*_{3l}}{2\varepsilon_0\sqrt{\int dV
    \frac{\partial (\omega\varepsilon)}{\partial \omega}E_{1i}^*E_{1i}
    \int dV \frac{\partial (\omega\varepsilon)}{\partial
      \omega}E_{3i}^*E_{3i}} \int dV \frac{\partial
    (\omega\varepsilon)}{\partial \omega}E_{2i}^*E_{2i}},
\label{eq:beta}
\end{align}
that depends on a spatial overlap of the linear cavity fields
$E_{\alpha i}$ over the (generally anisotropic) susceptibility
$\chi^{(3)}_{ijkl}$ of the nonlinear medium in space. The indices
$\alpha \in [1,2,3]$ and $\{i,j,k,l\} \in [x,y,z]$ run over the mode
number and cartesian components of the mode profiles, while
$\varepsilon_0$ and $\varepsilon$ denote the vacuum and relative
permitivity of the system.

Since typical pump energies tend to be much greater than the available
thermal energy in the system, i.e. $P \gg \gamma k_B T$, it is safe to
ignore the down-conversion term $-i\beta \omega_2 a_3 a_1^* a_2^*$ in
\eqref{a2}, in which case the equation for the pump decouples and is
linear in the incident drive field:
\[\frac{da_2}{dt}=(i\omega_2-\gamma_2)a_2 +
\sqrt{2\gamma_{2c}}s_{\text{in}}.
\]
Such an undepleted-pump approximation~\cite{rodriguez2007chi} greatly
simplifies the description of the four-wave mixing process, which is
thence described by the following coupled linear equations:
\begin{align}
  \frac{da_1}{dt} &= (i\omega_1-\gamma_1)a_1 - i\kappa e^{-2i\omega_2
    t} a_3 + \sqrt{2\gamma_{1d}}\xi_1 \label{eq:a1lin}
  \\\frac{da_3}{dt} &= (i\omega_3-\gamma_3)a_3 -
  i\frac{\omega_3}{\omega_1} \kappa^* e^{2i\omega_2 t} a_1 +
  \sqrt{2\gamma_{3d}}\xi_3,
\label{eq:a3lin}
\end{align}
Here, the role of the mediator mode at $\omega_2$ is captured by the
effectively linear coupling coefficient,
\begin{align}
\kappa&=\frac{2\beta\omega_1\gamma_{2c}P}{\gamma_2^2}.
\label{eq:nlcoupling}
\end{align}
Generally, in addition to frequency mixing, the Kerr nonlinearity also
leads to cross-phase modulation~\cite{boyd2003nonlinear}, which acts
to shift the resonator frequencies and hence disturbs frequency
matching,
$\omega_3=\omega_1+2\omega_2$~\cite{rodriguez2007chi,lin2014high}. Since
the corresponding modulation terms are temporally and hence
dynamically decoupled from the effective equations under the
undepleted approximation, in the following we account for such as well
as other possible sources of frequency mismatch, e.g. material
dispersion or even fabrication imperfections, by introducing a
time-independent frequency offset $\Delta\omega=
\omega_3-\omega_1-2\omega_2$ into one of the cavity frequencies.

To analyze energy transfer in this system, it suffices to consider the
rate of energy loss associated with each mode, $\frac{d|a_j|^2}{dt}$,
obtained through \eqref{a1lin} and \eqref{a3lin}. Collecting terms
proportional to the coupling coefficient $\kappa$, one finds that the
rate of energy extraction from $a_1$ is given by $\Pex= \langle
2\Im[\kappa^* \exp(2i\omega_2 t) a_3^* a_1]\rangle$, while the rate at
which energy is upconverted and gained by $a_3$ is given by
$\Pup=\langle 2\Im[\frac{\omega_3}{\omega_1}\kappa^* \exp(2i\omega_2
t) a_3^* a_1]\rangle = \frac{\omega_3}{\omega_1}\Pex$. Note that in
contrast to the case of two resonantly and linearly coupled
modes~\cite{otey2010thermal}, energy exchange within this four-wave
mixing process is not symmetric, i.e. $\Pex \neq \Pup$, but instead
satisfies a photon-number conservation
condition~\cite{ramirez2011degenerate}, i.e.
$\frac{\Pex}{\hbar\omega_1}=\frac{\Pup}{\hbar\omega_3}$, which ensures
that the number of photons lost by $a_1$ is equal to that gained by
$a_3$. Finally, the linearity of the coupled-mode equations allows the
extraction/upconversion rates and emission rates $P_{j} =
2\gamma_{jc}\langle|a_j|^2\rangle$ to be expressed in closed form,
leading to the following power spectral densities:
\begin{align}
\label{eq:pupc}
\Pex(\omega) &=\frac{4|\kappa|^2}{D_1(\omega)}
\left[-\gamma_1\gamma_{3d}\Theta(\omega+2\omega_2,T_3)+
  \gamma_{1d}\gamma_3(\omega_3/\omega_1)\Theta(\omega,T_1)\right]
\\ P_{1}(\omega) &= \frac{4\gamma_{1c}}{D_1(\omega)} \left[\gamma_{1d}
  |i(\omega-\omega_1-\Delta\omega)+\gamma_3|^2 \Theta(\omega,T_1) +
  \gamma_{3d}|\kappa|^2\Theta(\omega+2\omega_2,T_3)\right] \label{eq:p1out}
\\ P_{3}(\omega) &= \frac{4\gamma_{3c}}{D_3(\omega)} \left[\gamma_{3d}
  |i(\omega-\omega_3+\Delta\omega)+\gamma_1|^2 \Theta(\omega,T_3)+
  \gamma_{1d}(\omega_3/\omega_1)
  |\kappa|^2\Theta(\omega-2\omega_2,T_1)\right] \label{eq:p3out}
\end{align}
where $D_1(\omega) =
|(i(\omega-\omega_1)+\gamma_1)(i(\omega-\omega_1-\Delta\omega)
+\gamma_3)+\frac{\omega_3}{\omega_1} |\kappa|^2|^2$, $D_3=D_1(1
\leftrightarrow 3, \Delta\omega \rightarrow -\Delta\omega)$, and as
noted above, the upconverted power $\Pup=\frac{\omega_3}{\omega_1}
\Pex$.

The expressions above capture the most important features of this
four-wave mixing scheme. As an example, we consider the particular
scenario of equal-temperature $T$ cavities with
$\omega_1=k_\mathrm{B}T/\hbar$, $\omega_3=10\omega_1$, and zero
frequency mismatch $\Delta\omega=0$. We focus on the special situation
of resonances having equal dissipative and radiation rates,
$\gamma_{jd}=\gamma_{jc}=0.01\omega_1$, in which case both resonators
exhibit perfect thermal emissivities $\varphi_j \equiv
P_j/\Theta(\omega_j,T)$ (blue curves) in the absence of the pump,
i.e. $\kappa=0$. Note that we have chosen $\omega_3 \gg k_\mathrm{B}
T/\hbar = \omega_1$, in which case there is negligible radiation at
$\omega_3$ despite the near-unity emissivity.  \Figref{cmt}(b) shows
the thermal radiation spectrum near the two resonances for both
$\kappa=0$ (blue curve) and $\kappa = \gamma_1$ (red curves),
revealing giant enhancements in $\varphi_3 \gg 1$ in the presence of
the pump due to significant energy transfer from $\omega_1$ to
frequencies $\omega_3 \gg k_\mathrm{B}T/\hbar$ at which fluctuations
are otherwise exponentially suppressed by the Planck distribution
(black curve). \Figref{cmt}(c) shows $P_1(\omega)$ and $\Pex(\omega)$
near $\omega_1$, illustrating that the pump also causes the mode
resonances to split into Stokes ($-$) and anti-Stokes ($+$) peaks
which grow apart with increasing $\kappa$ and have center frequencies:
\begin{align*}
\omega_{1}^\pm &= \omega_1 + \frac{\Delta\omega \pm
  \sqrt{\Delta\omega^2+4(\gamma_1\gamma_3+\omega_3/\omega_1
    |\kappa|^2)}}{2} \\ \omega_{3}^\pm &= \omega_{1}^\pm(1
\leftrightarrow 3, \Delta\omega \rightarrow -\Delta\omega)
\end{align*}
As a consequence, in addition to mediating energy transfer, the
pump-induced red shift allows the resonator to effectively draw
additional energy available at longer wavelengths from the Planckian
reservoir, thus increasing its emission rate. Specifically, owing to
the red shift, the largest emissivity associated with the Stokes mode
is found to be $\varphi^\mathrm{max}_3 = k_\mathrm{B}
T/\Theta(\omega_3,T)$ rather than
$\Theta(\omega_1,T)/\Theta(\omega_3,T)$. Such enhancements, however,
are achieved only in the unrealistic limit of strong coupling $\kappa
\gg \gamma_{1,3}$, perfect frequency- and rate-matching,
$\Delta\omega=0$ and $\gamma_1=\gamma_3$, respectively, and negligible
spurious losses, $\gamma_{1c}=\gamma_{3d}=0$. The largest possible
extraction rate in this system, $\Pex^{\mathrm{max}} =
2\gamma_{1d}k_\mathrm{B}T$, occurs under similar conditions, except in
a regime wherein thermal excitations at $\omega_1$ are upconverted and
reabsorbed at $\omega_3$ at a faster rate than they decay in resonator
1, achieved in the limit of $|\kappa| \gg \gamma_3 \gg
\gamma_1$. While such a strong-coupling limit typically renders the
coupled-mode framework invalid~\cite{haus1984waves}, as we show below,
it is still possible to observe significant splitting in situations
$\kappa \gtrsim \gamma_j$ where coupled-mode theory is still
accurate. Moreover, while perfect frequency matching $\Delta\omega=0$
is desirable in order to guarantee optimal flux rates, in practice one
can still achieve significant transfer rates so long as the frequency
mismatch is smaller than either the resonator bandwidths,
$\Delta\omega \lesssim \gamma_j$, or the coupling rate, $\kappa
\gtrsim \gamma_j$. Finally, as expected, it follows from
\eqref{nlcoupling} that in order to achieve strong coupling
$|\kappa|$, one requires large nonlinear overlaps $\beta$, input power
$P$, and long lifetimes $1/\gamma_2$.

% Moreover, wider thermal bandwidths $\gamma_1,\gamma_3$ are
% preferable for increased upconversion and absorption rates unlike
% coherent frequency mixing applications that require small
% $\gamma_1,\gamma_3$ values for increased efficiencies.

The coupled-mode equations above describe a wide variety of triply
resonant systems, with the choice of implementation affecting only the
corresponding coupled-mode parameters. Aside from allowing different
temperature reservoirs, one reason to consider a system of three
physically distinct resonators is that it allows extraction of energy
from one of the resonators while mitigating additional heating
introduced by the coherent drive, thus paving the way for an active
near-field cooling mechanism analgous to a recently proposed scheme
based on gain media~\cite{ding2016active}. In what follows, we examine
a set of possible, physical implementations of this upconversion
scheme based on a configuration of two planar slabs supporting a
plurality of resonances and which allow significant extraction of
thermal energy from one slab to another.

\section{Near-field thermal upconversion and energy transfer between slabs}
\label{sec:RHT}

\begin{figure}[t!]
  \centering \includegraphics[width=1\linewidth]{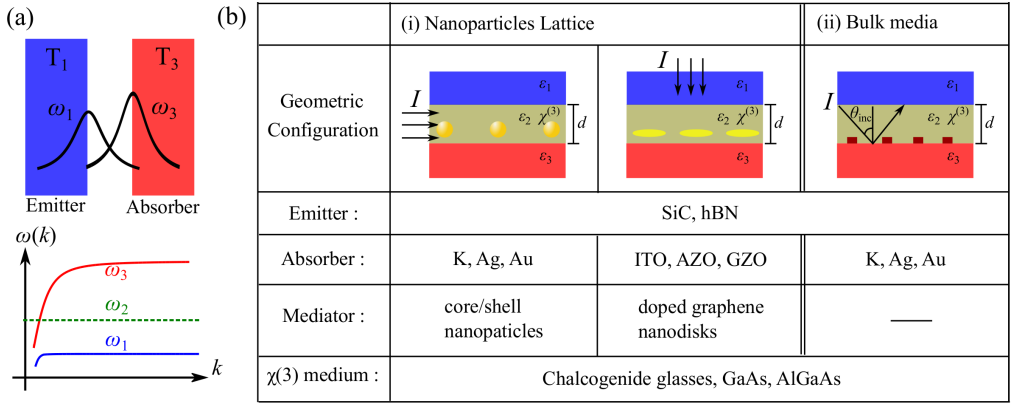}
  \caption{(a) Schematic of two plates consisting of different
    materials held at temperatures $T_1$ and $T_3$ and which support
    surface plasmon polaritons (SPPs) at far-apart resonant
    frequencies $\omega_1$ and $\omega_3$. Also shown is a schematic
    of the corresponding SPP dispersions $\omega_j(k)$ (blue and
    red). Thermal upconversion and transfer of energy from $\omega_1$
    to $\omega_3$ is facilitated by a mediator mode at $\omega_2 \sim
    \dfrac{(\omega_3-\omega_1)}{2}$ (green). (b) Table summarizing
    three different geometric configurations that result in
    significant thermal upconversion, along with various possible
    choices of emitting, absorbing, and nonlinear materials. The main
    difference between configurations is the choice of interveening
    medium, which consist of either a (i) lattice of nanoparticles
    embedded in the nonlinear medium or (ii) bulk nonlinear thin film,
    and serves as a tunable means to enhance the incident light.}
  \label{fig:scheme}
\end{figure}

Consider the planar geometry depicted schematically in
\figref{scheme}(a) and comprising two semi-infinite materials of
relative permittivities $\epsilon_1$ and $\epsilon_3$ which are held
at temperatures $T_1$ and $T_3$, respectively. The slabs are separated
by a gap of size $d$ that is filled with a $\chi^{(3)}$ nonlinear
medium of permittivity $\epsilon_2$. They support a large number of
SPPs localized around their respective interfaces and characterized by
their conserved in-plane momenta, $\vec{k}_{1}$ and $\vec{k}_3$, with
resonant frequencies $\omega_1(k_1)$ and $\omega_3(k_3)$ satisfying,
\begin{align}
e^{-2i\kappa_{z2}d}=\frac{(\epsilon_1\kappa_{z2}-\epsilon_2\kappa_{z1})
  (\epsilon_3\kappa_{z2}-\epsilon_2\kappa_{z3})}{(\epsilon_2\kappa_{z3}
  +\epsilon_3\kappa_{z2})(\epsilon_1\kappa_{z2}+\epsilon_2\kappa_{z1})}
\label{eq:dispersion}
\end{align}
where $k_{zj}=\sqrt{\epsilon_j\omega^2/c^2-k^2}$ and $k=|\vec{k}|$.
Such a configuration is known to result in large thermal energy
transfer when the two slabs are identical and hence support
equal-frequency resonances, $\omega_1 =
\omega_3$~\cite{francoeur2008near}. In what follows, we consider the
atypical situation of two different materials with dissimilar
dispersions and far-apart SPP resonance frequencies, where there is
negligible heat exchange under passive, non-equilibrium conditions but
where the presence of the nonlinear medium and a (mediator) mode at
$\omega_2 \approx \dfrac{\omega_3-\omega_1}{2}$ that is excited by
externally incident light can cause a large amount of thermal energy
in slab 1 (emitter) to be upconverted and absorbed in slab 3
(absorber). A schematic of the frequency dispersions of such a system
is shown in \figref{scheme}(a).  As discussed above, in order for any
two modes (in principles characterized by different $\vec{k}_{1,3}$)
to couple efficiently, their frequencies must satisfy the
frequency-matching condition $\Delta \omega \lesssim
\gamma_{1,3},\kappa$ described above. While material dispersion and
intrinsic fabrication imperfections generally preclude this from
occuring for all modes, such a condition can however be enforced by
appropiate nanostructuring and/or dispersion
engineering~\cite{zhao2011surface,pendry2004mimicking}. In principle,
while it is possible to achieve frequency matching for a plurality of
thermal modes by introducing and exciting multiple resonances near
$\omega_2$, such a situation would require a more complicated analysis
and is therefore out of the scope of this work.

For the sake of generality, we consider two classes of possible
geometric configurations, depicted in \figref{scheme}(b) and differing
primarily in the choice of material and interveening medium, which
involve either (i) nanostructured or (ii) bulk nonlinear materials
supporting SPP resonances at $\omega_2$. In (i), laterally incident
light couples to a periodic lattice of (composite nanospheres or
nanodisks) dipolar resonances while in (ii), vertically incident light
couples to a low-frequency SPP through a grating. As described below,
we ensure that regardless of implementation, an emitter mode at
$\vec{k}_1$ couples to a unique absorber mode at $\vec{k}_3$, which
greatly simplifies the calculation of flux transfer by avoiding
otherwise cumbersome analysis of multiply interacting degenerate
modes. Such a simplification also allows us to easily compute the
nonlinear coupling coefficient corresponding to each pair of modes
$\beta(\vec{k}_1,\vec{k}_3)$ and to exploit the analytical expression
for $\Pex(\omega,\vec{k}_1,\vec{k}_3)$ given in \eqref{pupc}. In
either configuration, the sparsity (low filling fractions) and
off-resonant nature of the nanoparticles and grating allow us to
ignore their impact on the translational symmetry and dispersion
relations of the slab modes.

The table in \figref{scheme}(b) summarizes a number of potential
configurations and material choices. For the sake of comparison and
consistency, we choose the emitter to be SiC, whose permittivity is
obtained from~\cite{palik1998handbook}, and the interveening nonlinear
medium to be ChG, whose permittivity $\epsilon_2=6.25$ and
$\chi^{(3)}\sim 1\times 10^{-17}$m$^2/V^2$ are taken from various
references~\cite{aio1978refractive,yan2011third,zakery2003optical,
  harbold2002highly}. For computational and conceptual convenience, we
assume an isotropic $\chi^{(3)}$ tensor with elements
$\chi_{xxxx}=3\chi_{xxyy}=3\chi_{xyxy}=\chi^{(3)}$~\cite{yan2011third,
  boyd2003nonlinear}. While there are many choices of possible
absorber materials, e.g. K, Ag, ITO, Au, and AZO, here we focus on a
select few, depending on the choice of implementation; their
permittivities are taken from several
references~\cite{west2010searching,franzen2008surface,blaber2009search}. Finally,
we fix the gap size to be $d=60$nm and consider only the situation in
which the entire system is at thermodynamic equilibrium, with
$T_1=T_3=300$K, for which there is zero heat exchange in the absence
of the pump.

\subsection{Nanoparticles lattice}
\label{sec:nanoparticles}

In this section, we examine situations in which the interveening
medium consists of a two-dimensional lattice of nanoparticles embedded
in ChG. The purpose of the lattice is threefold: First, it acts as a
grating which allows externally incident light to excite a given
(mediator) Bloch mode at $\omega_2$, described by the coupling
coefficient of \eqref{nlcoupling}, which couples SPPs at $\omega_1$
and $\omega_3$. Second, the particle shapes, sizes, and materials can
be exploited to engineer the resonant frequency $\omega_2$ and
associated decay rates, and thereby the pump field and coupling
coefficient $\kappa$, as needed. Third, the lattice provides many
degrees of freedom with which to enforce "quasiphase
matching"~\cite{bahabad2010quasi} over a broad range of $k$, allowing
the $\beta$ coefficient corresponding to multiple mode pairs
$(\vec{k}_1,\vec{k}_3)$ to be optimized.

Owing to the subwavelength size of the nanoparticles, their optical
response can be treated within a quasistatic, dipolar
approximation~\cite{bohren2008absorption}. They are also placed far
apart from one another, mitigating many-body scattering effects and
allowing the field induced by the incident drive to be conveniently
expressed as a linear superposition $\sum_p
E_2(\vec{x}_\parallel-\vec{x}_p,z)$ of the isolated particle
resonances, where $\vec{x}_\parallel$ and $\vec{x}_p$ denote the
transverse co-ordinates and center of each particle while
$E_2(\vec{x}_\parallel-\vec{x}_p,z)$ denotes its mode profile. The
$p$-polarized field profiles of the planar resonances are given by
$E_{jl}(z)e^{i\vec{k}_j.\vec{x}_{\parallel}}$, with $j=1,3$ and $l \in
\{x,y,z\}$ (nonzero components along ${\hat{\vec{k}}}$ and $\hat{z}$
directions), in which case the nonlinear coupling coefficient $\beta$
for a given mode pair is given by:
\begin{align}
  \beta(\vec{k}_1,\vec{k}_3) = \frac{\sum_p
    \int dV \chi_{ijk\ell}^{(3)} \,e^{i(\vec{k}_3-
      \vec{k}_1).\vec{x}_{\parallel}}
    E_{1i}(z)E_{2j}(\vec{x}_{\parallel}-\vec{x}_p,z)
    E_{2k}(\vec{x}_{\parallel}-\vec{x}_p,z)
    E_{3\ell}^*(z)}{2\epsilon_0 (\sum_p \int dV \frac{\partial
      \epsilon\omega}{\partial\omega}|E_{2}(\vec{x}_\parallel-\vec{x}_p,z)|^2)
    (\int dV \frac{\partial
      \epsilon\omega}{\partial\omega}|E_{1}(z)|^2 )^{1/2} (\int dV
    \frac{\partial \epsilon\omega}{\partial\omega}|E_{3}(z)|^2 )^{1/2}
  }
\label{eq:betaplates}
\end{align}
Here, the sum is taken with respect to the index $p$ of each particle,
$\{i,j,k,\ell\}$ denote cartesian field components, and
$\chi^{(3)}_{ijk\ell}$ is the Kerr tensor of the background
medium. The SPP dispersions $\omega_j(k)$ and corresponding (purely
disspative) decay rates $\gamma_{jd}(k)$ are obtained by the
complex-frequency solutions of~\eqref{dispersion}. Because of the
relatively low in-plane and volume filling fractions as well as the
off-resonant response of the nanoparticle lattice, they are assumed to
have a negligible effect on the planar dispersion and mode profiles.

In order to take advantage of the large density of states available in
the near field, we choose geometric and material parameters that
ensure large $\beta$ for as wide a range of modes as possible.  We
restrict our analysis to modes having equal momentum $\vec{k} =
\vec{k}_1=\vec{k}_3$, in which case the integral in \eqref{betaplates}
no longer exhibits a phase factor that depends on
$\vec{k}_{1,3}$. Such a simplifying assumption is further justified by
the fact that although technically the nonlinearity can couple modes
propagating in different directions ($\vec{k}_1 \neq \vec{k}_3$), a
situation that arises in the configuration of \secref{bulk}, these
typically suffer from diminished mode overlaps and hence reduce the
overall upconversion efficiency. Our analysis is further simplified by
ensuring that the induced particle fields exhibit cylindrical symmetry
about the $\hat{z}$ axis, in which case the field
$E_2(\vec{x}_\parallel-\vec{x}_p)$ and hence the coupling between each
mode pair is independent of the $\hat{\vec{k}}$ direction. The chosen
implementions involve either a lattice of spherical nanoparticles or
graphene nanodisks, respectively, which act like $\hat{z}$-oriented
dipoles under illumination by light incident along either the parallel
or $\hat{z}$ directions, respectively. These considerations allow us
to write $\Pex(\omega,\vec{k}_1,\vec{k}_3)$ as $\Pex(\omega,k)$, and
thus to express the net heat extraction rate per unit area $H$ across
the gap as:
\begin{align}
  H=\int_0^\infty \frac{d\omega}{2\pi} \underbrace{\int_0^{\infty}
    \frac{dk}{2\pi} k \Pex(\omega,k)}_{\Phi_\mathrm{ex}(\omega)}
\label{eq:Phi}
\end{align}
where we further define the quantity $\Pex(k)
=\int\frac{d\omega}{2\pi} \Pex(\omega,k)$ as the frequency-integrated
flux at each $k$. Similarly, one can define the associated flux
spectral density, $\Phi_{\text{ex}}(\omega)=\int_0^{\infty}
\frac{dk}{2\pi} k \Pex(\omega,k)$, by integrating instead over all
wavevectors.  Finally, in addition to the spatial profile of each mode
pair, it is equally important to enforce the frequency-matching
condition $\Delta\omega(k)=\omega_3(k)- \omega_1(k)-2\omega_2 \lesssim
\gamma_j,\kappa$, which is generally not guaranteed and as illustrated
below, depends sensitively on the choice of materials.

\begin{figure}[t!]
  \centering \includegraphics[width=1\linewidth]{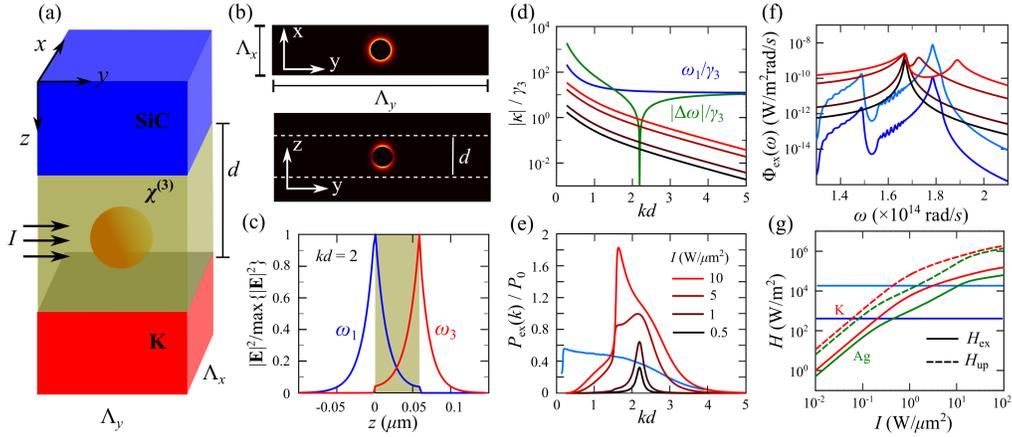}
  \caption{(a) Schematic of a planar system of SiC and K slabs at
    thermal equilibrium (room temperature) separated by a gap of size
    $d=60$nm that is filled with a rectangular lattice of nanospheres
    with unit-cell size $\Lambda_x \times \Lambda_y$ embedded in a
    $\chi^{(3)}$ nonlinear medium (ChG). The SiO$_2$/Au core/shell
    nanoparticles have core/shell radii of 15nm and 20nm,
    respectively, and the dimensions of the unit cell are
    $\Lambda_x=5R$ and $\Lambda_y=\pi c/\omega_2$. The SiC and K slabs
    support SPPs at frequencies $\omega_1$ and $\omega_3$,
    respectively, which couple nonlinearly to dipolar particle
    resonances at $\omega_2 \sim \frac{1}{2}(\omega_3-\omega_1)$
    excited by a laterally incident, monochromatic,
    $\hat{z}$-polarized planewave of frequency $\omega_2$ and
    intensity $I$. (b) $xy$ and $yz$ cross sections of the particle
    resonances $|\vec{E}_2|^2$, with yellow/black denoting
    maximum/zero amplitude. (c) Normalized slab mode-profiles at a
    representative wavenumber $k d=2$, with the shaded region
    indicating the interveening medium.  (d) Variations in the
    nonlinear coupling $\kappa$, frequency mismatch $\Delta\omega$,
    and frequency $\omega_1$ with respect to $kd$, normalized by the
    corresponding disspation rate $\gamma_3$ of the K slab. (e)
    Frequency-integrated heat-extraction spectrum $\Pex(k)$,
    normalized by $P_0=2\gamma_1\Theta(\omega_1,T)$, as a function of
    $k d$ and for multiple incident intensities $I$. Also shown are
    the (f) associated spectral densitiy $\Phi_{\text{ex}}(\omega)$
    and (g) net extracted and upconverted flux rates, $H_{\text{ex}}$
    and $H_{\text{up}}$, as a function of $I$. For comparison, (e--g)
    also show the heat-transfer rates associated with two
    vacuum-separated SiC slabs held at either 300K (light blue) or 1K
    (dark blue) temperature differences.}
  \label{fig:sphere}
\end{figure}

\subsubsection{Nanospheres}
\label{sec:nanospheres}

\Figref{sphere}(a) shows a rectangular lattice of nanospheres of equal
radii $R$ and unit-cell size $\Lambda_x \times \Lambda_y$ that is
illuminated by a $z$-polarized incident wave $\propto \hat{z}
e^{-i(k_2 y+\omega_2 t)}$, with
$k_2=\sqrt{\varepsilon_2}\omega_2/c$. The incident field excites
dipolar modes with field profiles of the form
$E_{2j}(\vec{x}_\parallel-\vec{x}_p,z)e^{ik_2 y}$. Since for any given
mode pair the nonlinear overlap integral in \eqref{betaplates}
involves a sum over all particle positions, vanishing unless the
individual unit-cell contributions add constructively, we make an
appropiate choice of period $\Lambda_y=\pi/k_2$. Furthermore, although
the lack of a ``Bloch phase'' in the $\hat{\vec{x}}$ direction means
that at least in principle $\Lambda_x$ can be much smaller than
$\Lambda_y$, here we choose a large enough $\Lambda_x=5R$ in order to
ignore many-body effects on the induced field. Taking advantage of the
equal contribution of each unit cell to \eqref{betaplates} and
restricting our analysis to momentum-matched mode pairs
$\vec{k}_1=\vec{k}_3$, the coupling $\kappa(k)$ is given by:
\begin{align}
\kappa(k)= \frac{\omega_1\sigma_{\text{abs}}I
  \int_{\text{cell}}\chi_{ijk\ell}^{(3)}E_{1i}(z)
  E_{2j}(\vec{x}_\parallel-\vec{x}_p,z)
  E_{2k}(\vec{x}_\parallel-\vec{x}_p,z)E_{3\ell}^{*}(z)}{4\epsilon_0
  \gamma_{2d}\Lambda_x \Lambda_y (\int_{\text{cell}} \frac{\partial
    \epsilon\omega}{\partial
    \omega}|E_2(\vec{x}_\parallel-\vec{x}_p,z)|^2)(\int dz
  \frac{\partial \epsilon\omega}{\partial\omega}|E_1(z)|^2)^{1/2}(\int
  dz \frac{\partial\epsilon\omega}{\partial\omega}|E_3(z)|^2)^{1/2}}
\label{eq:kappakp}
\end{align}
where $\omega_2$, $\sigma_{\text{abs}}$, and $\gamma_{2d}$ are the
resonance frequency, absorption cross-section, and dissipation rate of
the nanoparticle resonances, and $I$ denotes the incident-field
intensity. To ensure frequency matching, we employ core-shell
nanoparticles~\cite{ghosh2011core,oldenburg1998nano} with carefully
chosen core/shell radii and material
dispersions~\cite{sihvola2006peculiarities}. In particular, we choose
particles made of silica~\cite{kitamura2007optical} core and
Au~\cite{west2010searching} shells having inner and outer radii of
$15$nm and $20$nm, respectively. These parameters yield a resonance
frequency $\omega_2=8.74\times 10^{14}$rad/s, dissipative and
radiative decay rates, $\gamma_{2d}=4.31\times 10^{12}$rad/s and
$\gamma_{2c}=8.29\times 10^{12}$rad/s, respectively, and lattice
parameters $\Lambda_x=5R=100$nm and $\Lambda_y=\pi c/\omega_2=431$nm,
resulting in volume and in-plane filling fractions of 0.01 and 0.03,
respectively. The corresponding absorption cross-section
$\sigma_{\text{abs}}=1.2\times 10^{-13}$m$^{2}$ can also be obtained
via a well-known dipolar
analysis~\cite{bohren2008absorption,sihvola2006peculiarities}.
Choosing SiC as the emitter and K as the absorber means that the two
interacting slab resonances have frequencies $\omega_1 \sim 1.68\times
10^{14}$rad/s ($\lambda_1 \sim 11.22\mu$m) and $\omega_3 \sim 2\times
10^{15}$rad/s ($\lambda_3 \sim 0.94\mu$m), respectively.

\Figref{sphere}(b) shows the $xy$ and $yz$ cross-sections of the
nanosphere mode profiles $|\vec{E}|^2$ within a unit cell, where
black/yellow represent zero/maximum values. The spatial mode profiles
of the slab resonances are illustrated in (c) at a representative $k
d=2$, where the shaded region indicates the interveening
medium. \Figref{sphere}(d) shows the variation in the coupling
coefficient $|\kappa|$ for increasing intensities $I = \{0.5, 1, 5,
10\}~\mathrm{W}/\mu\mathrm{m}^2$ (from black to red), frequency
mismatch $\Delta \omega$, and resonance frequency $\omega_1$, as a
function of the dimensionless wavenumber $k d$. All quantities are
normalized by the corresponding dissipation rate $\gamma_3$ at each
$k$, which is the largest of the loss rates in the
system. \Figref{sphere}(e) shows the ratio of $\Pex(k)$ to the thermal
radiation rate of an isolated thermal resonance
$P_0=2\gamma_1\Theta(\omega_1,T_1)$ as a function of $k d$, while
\figref{sphere}(f) shows the associated spectrum
$\Phi_{\text{ex}}(\omega)$, in units of~$W/\mathrm{m}^2 s$, for
multiple $I$. Finally, the integrated extraction $H_{\text{ex}}$ and
upconversion $H_{\text{up}}$ rates are plotted in \figref{sphere}(g)
as a function of $I$.

We now explain the most salient and important features associated with
heat exchange in this system. As shown in (e), $\Pex(k)$ exhibits a
small peak at $k d \approx 2$ that grows and widens with increasing
$I$, causing the overall heat transfer per unit area $H_{\text{ex}}$
to monotonically increase and eventually saturate as $I \to
\infty$. The associated spectrum $\Phi_{\text{ex}}(\omega)$ also shows
a corresponding increase along with linewidth broadening with
increasing $I$. These features are explained as follows: First,
\eqref{pupc} shows that in the weak coupling regime $|\kappa|/\gamma_3
\ll 1$, the flux rate $H_{\text{ex}} \sim |\kappa|^2 \sim
I^2$. Second, as evident from \figref{sphere}(d), the bandwidth of the
spectrum is primarily determined by the range of modes satisfying the
frequency matching constraint, $\Delta\omega \lesssim
\gamma_j$,$|\kappa|$, with the peak occurring at the value of $k$
which minimizes $\Delta \omega$. At small $I$ or equivalently,
$\kappa/\gamma_3 \ll 1$, the range of modes satisfying $\Delta\omega
\lesssim \gamma_3$ is narrow, but higher intensities and hence
increasing $\kappa$ allow frequency matching to be satisfied over a
wider range of $k$. Third, the rapid decrease in the flux rate at
large $k d \gg 1$ happens because $\beta$ and hence $\kappa$ depend on
the spatial decay of the planar mode profiles, which become
increasingly localized to the surface with increasing $k$. The precise
value of $k$ at which $\Delta\omega(k)=0$ depends primarily on the
choice of materials and geometric parameters: although its occurence
at larger $k$ would facilitate increased bandwidths due to the lack of
dispersion at large $k \gg \omega/c$, the resulting exponential
suppression in $\beta$ suggests instead an optimal choice of particle
parameters for a given $d$, here chosen at an intermediate $k d
\approx 2$. Note also that we only consider $\Delta\omega$ arising
from material dispersion and ignore power-dependent shifts due to
cross-phase modulation,~\cite{rodriguez2007chi} which tend to be small
and in any case can be compensated by suitable lattice
parameters. Finally, as discussed above, the ratio $\Pex/P_0$ can
exceed one when the system enters the strong coupling regime $\gamma_3
\lesssim |\kappa| \ll \omega_1$, in which case the Stokes peak can
draw additional thermal energy available at longer wavelengths.

\Figref{sphere} also shows the flux rate associated with a symmetric
configuration of two identical SiC plates maintained at $T_1$ and
$T_3$ and separated by the same gap size (albeit in vacuum). The net
heat-transfer rate in this more typical scenario can be computed via
the well-known fluctuational electrodynamics
framework~\cite{basu2009review} and is given by:
\begin{align}
  H^{(0)} = \int_0^{\infty} \frac{d\omega}{2\pi}
  \left[\Theta(\omega,T_1)-\Theta(\omega,T_3)\right]
  \underbrace{\int_0^\infty dk k
    \frac{2\Im\{r_{21}^p\}\Im\{r_{23}^p\}e^{-2\Im\{k_{z2}\}d}}{
      \pi|1-r_{21}^p r_{23}^p e^{-2\Im\{k_{z2}\}d}|^2}
  }_{\Phi^{(0)}(\omega)}
\label{Pklin}
\end{align}
where $r_{j\ell}^p$ is the $p$-polarized Fresnel reflection
coefficient between mediums $j$ and $\ell$ and where, as before, one
can define and compute the frequency-integrated flux $P^{(0)}(k)$ and
flux spectral density $\Phi^{(0)}(\omega)$. For the sake of
comparison, we consider two different operating conditions
corresponding to either (i) large ($T_1=300$~K, $T_3=0$~K) (light blue
curves) or (ii) small ($T_1=301$~K, $T_3=300$~K) (dark blue curves)
temperature differentials.

From \figref{sphere}(e), one immediately observes that compared to the
nonlinear scheme, the exponential decay in $\Pex^{(0)}(k)$ occurs at
smaller $k d\approx 1$. The reasons are twofold: First, the increased
proximity of the nanoparticle resonances to the slab interfaces
results in slightly larger mode overlaps. Second and most importantly,
increasing $I$ and hence $\kappa$ allows thermal energy in the SiC to
be upconverted and then absorbed at a much faster rate than it is
dissipated in the SiC. Consequently, the range of participating modes
and hence the bandwidth of $\Pex(k)$ increases with increasing
$I$. When combined with the aforementioned Stokes enhancement, the net
effect is a significant increase in the nonlinear flux rates, which
exceed the corresponding linear flux rates $\approx
10^4~\mathrm{W}/\mathrm{m}^2$ at a temperature-dependent threshold
intensity $I_c$, which is approximately $I_c \approx
3\mathrm{W}/\mu\mathrm{m}^2$ in scenario (i) and $I_c \approx
0.1\mathrm{W}/\mu\mathrm{m}^2$ in scenario (ii). Finally, we note that
in contrast to the typically investigated system of two
vacuum-separated identical slabs, the frequency-matching wavenumber
$\kappa$ and hence peak value of $k$ in the nonlinear scheme can be
tuned by engineering the core/shell nanoparticle geometry or by an
appropiate choice of emitter/absorber materials. Nevertheless, we find
that the threshold intensities do not vary much with respect to the
choice of absorber material. As shown in \figref{sphere}(f) (green
curves), fixing the parameters of the emitter and core/shell
nanoparticles but replacing K with Ag also leads to significant energy
transfer (albeit slighly smaller owing to the smaller
frequency-matching wavenumber).

Finally, we remark that our choice of geometry and operating
conditions is by no means optimal. For instance, one could further
increase the mode overlaps and frequency-matching wavenumbers by
exploiting more complicated and diverse particle shapes (e.g. nanorods
or even asymmetric particles) or by employing other emitter materials
(e.g. hBN, Au), leading to greater efficiencies and lower power
requirements. Moreover, while the choice of lattice period in this
example is primarily motivated by the need to enforce quasiphase
matching for a laterally incident pump, as we show in the next
section, that and other geometric considerations are also highly
dependent on the choice of incident direction and/or particle shape.

\begin{figure}[t!]
  \centering
  \includegraphics[width=1\linewidth]{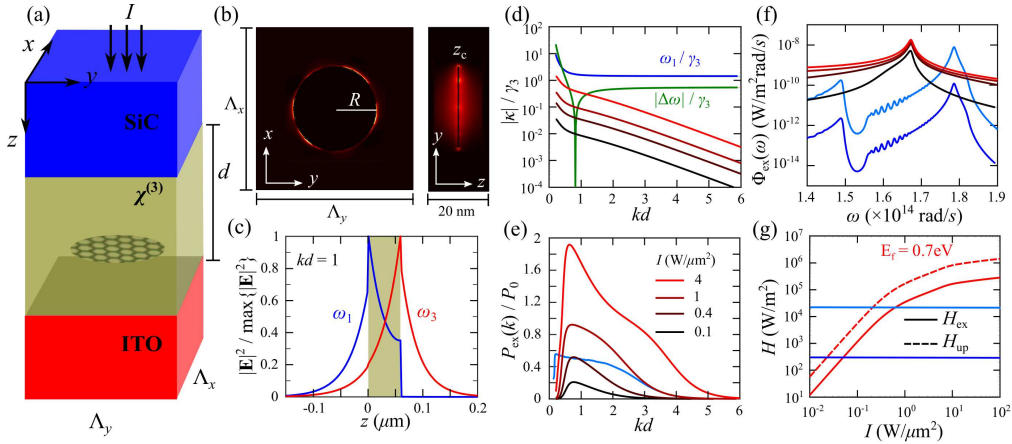}
  \caption{(a) Schematic of a planar system of SiC and ITO slabs at
    thermal equilibrium (room temperature) separated by a gap of size
    $d=60$nm that is filled with a square lattice of doped graphene
    nanodisks with unit-cell size $\Lambda \times \Lambda $ embedded
    in a $\chi^{(3)}$ nonlinear medium (ChG). The graphene nanodisks
    have radius $R=20$nm, Fermi energy $E_\mathrm{F}=0.7eV$ and are
    placed at a distance of $10$nm from ITO slab. The SiC and ITO
    slabs support SPPs at frequencies $\omega_1$ and $\omega_3$,
    respectively, which couple nonlinearly to nanodisk resonances at
    $\omega_2 \sim \frac{1}{2}(\omega_3-\omega_1)$ excited by a
    normally incident, monochromatic, $\hat{x}+i\hat{y}$-polarized
    planewave of frequency $\omega_2$ and intensity $I$. (b) $xy$ and
    $yz$ cross sections of the particle resonances $|\vec{E}_2|^2$,
    with yellow/black denoting maximum/zero amplitude. (c) Normalized
    slab mode-profiles at a representative wavenumber $k d=1$, with
    the shaded region indicating the interveening medium.  (d)
    Variations in the nonlinear coupling $\kappa$, frequency mismatch
    $\Delta\omega$, and frequency $\omega_1$ with respect to $kd$,
    normalized by the corresponding disspation rate $\gamma_3$ of the
    ITO slab. (e) Frequency-integrated heat-extraction spectrum
    $\Pex(k)$, normalized by $P_0=2\gamma_1\Theta(\omega_1,T)$, as a
    function of $k d$ and for multiple incident intensities $I$. Also
    shown are the (f) associated spectral densitiy
    $\Phi_{\text{ex}}(\omega)$ and (g) net extracted and upconverted
    flux rates, $H_{\text{ex}}$ and $H_{\text{up}}$, as a function of
    $I$. For comparison, (e--g) also show the heat-transfer rates
    associated with two vacuum-separated SiC slabs held at either 300K
    (light blue) or 1K (dark blue) temperature differences.}
  \label{fig:graphene}
\end{figure}

\subsubsection{Nanodisks} 
\label{sec:nanodisks}

\Figref{graphene}(a) shows a square lattice of doped graphene
nanodisks of unit-cell size $\Lambda \times \Lambda$ that is
illuminated by a normally incident, circularly polarized wave $\propto
(\hat{x}+i\hat{y}) e^{-i(k_2 z+\omega_2 t)}$, with
$k_2=\sqrt{\varepsilon_2}\omega_2/c$.  Owing to the normal incidence,
the induced fields $E_{2j}(\vec{x}_\parallel-\vec{x}_p,z)$ at each
lattice site have the same phase and consequently, each individual
unit cell contributes equally irrespective of the lattice period. For
simplicity, we choose $\Lambda=3R$ which allows us to ignore many-body
scattering. As before, the cylindrical symmetry of the disks and the
circular polarization of the incident light imply that
$\beta(\vec{k}_1,\vec{k}_2) \to \beta(k)$ is independent of the
$\hat{\vec{k}}$ direction. Our choice of graphene as opposed to other
potential polaritonic/plasmonic materials (e.g. Au, Ag, etc) is
motivated by the large degree of tunability in the resonance frequency
$\omega_2$ and hence frequency-matching wavenumber with respect to the
Fermi energy or doping concentration of graphene, as well as by recent
experiments exploring related
structures~\cite{fang2013gated,koppens2011graphene}. In what follows,
we choose graphene nanodisks of radius $R=20$nm, fermi energy
$E_f=0.7$eV, and intrinsic lifetime $\tau=6\times 10^{-13}$,
consistent with recent experimental
measurements~\cite{koppens2011graphene}, which results in resonance
frequencies $\omega_2=3.04\times 10^{14}$rad/s and dissipation rates
$\gamma_{2d}=3.3\times 10^{12}$rad/s. For an incident light of
intensity $I$, the coupling $\kappa(k)$ is given by \eqref{kappakp},
where $\sigma_{abs}=2\times 10^{-14}$m$^{2}$ is calculated from the
effective dipolar susceptibility~\cite{fang2013gated}. The
corresponding field profiles are obtained via numerical simulation of
Maxwell's equations using the boundary element method
(BEM)~\cite{hohenester2012mnpbem}, where the graphene nanodisk is
assumed to have an effective thickness $h=0.5$nm and dielectric
permittivity $\epsilon_G=1-\dfrac{4\pi i \sigma(\omega)}{\omega h}$,
with $\sigma(\omega)$ denoting the sheet conductivity of
graphene~\cite{fang2013gated}. Assuming SiC and ITO as the emitter and
absorber in this configuration, respectively, one finds $\omega_1 \sim
1.68\times 10^{14}$rad/s ($\lambda_1 \sim 11.22\mu$m) and $\omega_3
\sim 8.4\times 10^{14}$rad/s ($\lambda_3 \sim 2.24\mu$m).

\Figref{graphene}(b) shows the $xy$ and $yz$ cross-sections of the
nanodisk mode profiles $|\vec{E}|^2$ within a unit cell. We remark
that since the modes are largely confined to the nanodisk
circumference, further close packing or potentially smaller lattice
periods could potentially be employed to enhance
$\kappa$. \Figref{graphene}(c) shows profiles of the planar resonances
at a representative $kd=1$, with the shaded region indicating the
interveening medium. While most of the features observed in this
configuration and illustrated in \figref{graphene}(d--g) are
qualitatively similar to the previous implementation in
\secref{nanospheres} based on nanospheres, we emphasize some of the
main differences. First, the possibility of exciting cylindrically
symmetric nanoparticle resonances with normally incident light allows
a reduction of the lattice period and leads to slighly reduced
threshold intensities, with $I_c \approx 0.5
\mathrm{W}/\mu\mathrm{m}^2$ in the scenario corresponding to SiC slabs
held at a 300K temperature difference (light blue curves). Second,
graphene offers additional (potentially dynamic) tunability with
respect to the frequency-matching wavenumber, leading to large (order
of magnitude) differences in the heat flux for relatively small
changes in $E_f$ (not shown). Third, the two-dimensional nature of
graphene nanodisks allows greater choice in their vertical placement;
one could for instance consider mutliple sets of nanodisks distributed
vertically throughout the interveening medium, which would further
increase $\kappa$. One important constraint to consider when choosing
the shapes of the nanoparticles is the choice of pump and wavelengths;
for instance, while graphene supports polaritons in the mid-infrared,
other metallic nanoparticles can support resonances at higher
(e.g. near-infrared or visible) wavelengths. Finally, we note that our
choice of direction for the incident pump in either scenario is
motivated by the desire to restrict our analysis to cylindrically
symmetric nonlinear coupling coefficients (i.e. independent of the SPP
propagation direction). One could also tune the lattice parameters and
particle shapes so as to allow obliquely incident pumps, though this
would likely lead to much more compliated
$\beta(\vec{k}_{1},\vec{k}_{3})$. We analyze such a situation in the
following section, in which we study upconversion in the case of a
bulk nonlinear thin film as the interveening medium.

\subsection{Bulk media}
\label{sec:bulk}

\begin{figure}[t!]
  \centering \includegraphics[width=1\linewidth]{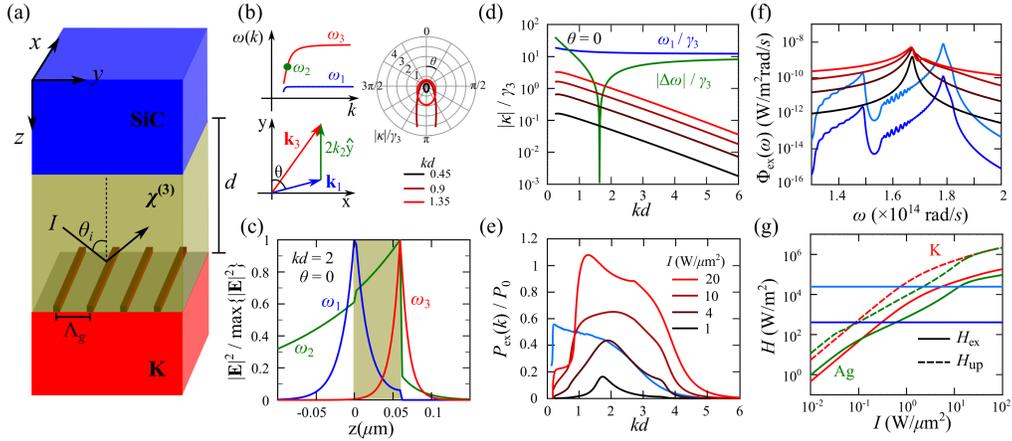}
  \caption{(a) Schematic of a planar system of SiC and K slabs at
    thermal equilibrium (room temperature) separated by a gap of size
    $d=60$nm that is filled with $\chi^{(3)}$ nonlinear medium
    (ChG). The SiC and K slabs support SPPs at frequencies $\omega_1$
    and $\omega_3$, respectively, which couple nonlinearly to a SPP
    resonance at $\omega_2 \sim \frac{1}{2}(\omega_3-\omega_1)$ that
    is excited via a grating (of period $\Lambda$) by monochromatic
    light of frequency $\omega_2$, intensity $I$, and wavevector
    $k_{2y}$ incident at an angle $\theta_\mathrm{inc}$ with respect
    to the $\hat{z}$ direction. (b) Schematic illustation of the
    coupling of SPP resonances of wavevectors $\vec{k}_{1}$ and
    $\vec{k}_3$; the polar plot shows the directional dependence of
    the nonlinear coupling coefficient $\kappa(k,\theta)$, where
    $\vec{k}_{1}=(k,\theta)$ is expressed in polar coordinates, with
    $\theta$ denoting the angle extended by $\vec{k}_1$ with respect
    to $\hat{y}$.  (c) Normalized slab mode-profiles at a
    representative wavenumber $k d=2$ and angle $\theta=0$, with the
    shaded region indicating the interveening medium.  (d) Variations
    in the nonlinear coupling $\kappa$, frequency mismatch
    $\Delta\omega$, and frequency $\omega_1$ with respect to $kd$,
    normalized by the corresponding disspation rate $\gamma_3$ of the
    K slab. (e) Frequency-integrated, angle-averaged heat-extraction
    spectrum $\Pex(k)$, normalized by
    $P_0=2\gamma_1\Theta(\omega_1,T)$, as a function of $k d$ and for
    multiple incident intensities $I$. Also shown are the (f)
    associated spectral densitiy $\Phi_{\text{ex}}(\omega)$ and (g)
    net extracted and upconverted flux rates, $H_{\text{ex}}$ and
    $H_{\text{up}}$, as a function of $I$. For comparison, (e--g) show
    the heat-transfer rates associated with two vacuum-separated SiC
    slabs held at either 300K (light blue) or 1K (dark blue)
    temperature differences.}
  \label{fig:plates}
\end{figure}

\Figref{plates}(a) shows a 1d $y$-periodic grating of period $\Lambda$
resting at the interface of the ChG and K slabs. The grating is
assumed to have a negligible effect on the SPPs at $\omega_1$ and
$\omega_3$, and chosen so that incident light of wavevector $k_{2y}$
along the $y$ direction can excite a SPP of frequency
$\omega_2=\omega_3(k_{2y})$ localized around the ChG--K interface. The
angle of incidence $\theta_\mathrm{inc}$ with respect to the $\hat{z}$
axis is chosen so as to satisfy,
\[
\dfrac{\epsilon_2\omega_2\sin(\theta_\mathrm{inc})}{c}+\dfrac{2\pi}{\Lambda}
=k_{2y},\] thus ensuring that only the first diffracted order is
excited by the
pump~\cite{sambles1991optical,porto1999transmission}. The
$p$-polarized field profiles of the planar resonances are given by
$E_{j\ell}(z)e^{i\vec{k}_j.\vec{x}_{\parallel}}$, with $j\in [1,3]$
and $\ell \in \{x,y,z\}$ (nonzero components along ${\hat{\vec{k}}}$
and $\hat{z}$), in which case the nonlinear coupling $\beta$ is given
by:
\begin{align}
  \beta(\vec{k}_1,\vec{k}_3) = \frac{ \int dV
    \chi_{ijk\ell}^{(3)} \,e^{i(\vec{k}_3-
      \vec{k}_1-2k_{2y}\hat{y}).\vec{x}_{\parallel}}
    E_{1i}(z)E_{2j}(\omega_2,z)E_{2k}(\omega_2,z)
    E_{3\ell}^*(z)}{2\epsilon_0 (\int dV \frac{\partial
      \epsilon\omega}{\partial\omega}|E_{2}(\omega_2,z)|^2) (\int dV
    \frac{\partial \epsilon\omega}{\partial\omega}|E_{1}(z)|^2 )^{1/2}
    (\int dV \frac{\partial
      \epsilon\omega}{\partial\omega}|E_{3}(z)|^2 )^{1/2} }
\label{eq:betaplates2}
\end{align}
Notably, one finds that nonzero coupling is only possible under the
momentum-matching condition, $\vec{k}_1+2k_{2y}\hat{y}=\vec{k}_3$,
represented schematically in (b). Such a condition ensures that a wave
at $\vec{k}_1$ couples to a unique wave at $\vec{k}_3$, in which case
$\beta(\vec{k}_1,\vec{k}_3) \to \beta(k,\theta)$, where $k =
|\vec{k}_1|$ and $\theta$ is the angle extended by $\vec{k}_1$ with
respect to the $y$ axis. Employing the translational invariance of the
overlap integrals for such a process, one finds that coupling
$\kappa(k,\theta)$ is given by:
\begin{align}
  \kappa(k,\theta) = \frac{\omega_1\gamma_{2c}I\int
    dz\chi_{ijk\ell}^{(3)} E_{1i}(\omega_1,z) E_{2j}(\omega_2,z)
    E_{2k}(\omega_2,z) E_{3\ell}^*(\omega_3,z)}{\epsilon_0\gamma_2^2
    \int dz\frac{\partial
      \epsilon\omega}{\partial\omega}|E_{2}(\omega_2,z)|^2 (\int dz
    \frac{\partial \epsilon\omega}{\partial
      \omega}|E_{1}(\omega_1,z)|^2)^{1/2} (\int dz
    \frac{\partial\epsilon\omega}{\partial\omega}
    |E_{3}(\omega_3,z)|^2)^{1/2}}
\end{align}
where as before, $I$ is the intensity of the incident light and
$\gamma_{2c}$ is obtained by solving the full scattering
problem~\cite{sambles1991optical,porto1999transmission}. The net
heat-transfer rate per unit area across the gap is then given by
$H_{\text{ex}}(k)=\int_0^{2\pi}\frac{d\theta}{2\pi}
\int_0^{\infty}\frac{d\omega}{2\pi} \Pex(\omega,k,\theta)$, where
$\Pex$ follows from \eqref{Phi} and the associated flux spectral
density is given by
$\Phi_{\text{ex}}(\omega)=\int_0^{2\pi}\frac{d\theta}{2\pi}\int_0^{\infty}
\frac{dk}{2\pi} k \Pex(\omega,k,\theta)$.  In what follows, we choose
a grating period $\Lambda=205$nm in order to couple incident light at
an angle $\theta_\mathrm{inc}=\pi/4$ to a SPP of frequency
$\omega_2=9\times 10^{14}$rad/s, wavenumber $k_{2y}=2.8\omega_2/c$,
and dissipation rate $\gamma_{2d}=2.3\times 10^{12}$rad/s. For
convenience and expediency, we ignore the impact of the grating on the
dispersions at $\omega_{1,3}$ and assume critical coupling,
$\gamma_{2c}=\gamma_{2d}$. Of course, any deviation from this
condition would result in decreased
efficiency~\cite{joannopoulos2011photonic}. Choosing SiC and K as the
emitter and absorber materials, respectively, one finds that $\omega_1
\sim 1.68\times 10^{14}$rad/s ($\lambda_1\sim 11.22\mu$m) and $\omega_3
\sim 2\times 10^{15}$rad/s ($\lambda_2\sim 0.94\mu$m).

\Figref{plates}(c) shows the various mode profiles $|\vec{E}|^2$ at a
representative $k d=2$ and for $\theta=0$, illustrating the larger
spatial extent of the mediator mode owing to its effectively smaller
wavenumber $k_{2y}d\approx 0.5$. The polar plot in \figref{plates}(b)
shows the normalized $|\kappa(k,\theta)|/\gamma_3$ for increasing
values of $k d$, illustrating the directional dependence of the
coupling coefficient and the lack of frequency matching stemming from
material dispersion and the absence of a $\omega_3(\vec{k}_3)$ mode in
certain directions.

\Figref{plates}(d,f,g) shows that heat exchange in this configuration
is qualtiatively similar to what is observed in the case of
nanoparticle lattices. Despite the similar frequency spectrum and net
flux rate dependence on $I$, the complicated dependence of $\kappa$ on
both $k$ and $\theta$ does lead to noticeably different $\Pex(k)$,
shown in (e). Moreover, while this configuration has the advantage of
simplicity and ease of realization, as expected the significantly
decreased modal overlaps lead to much larger power requirements. For
instance, the threshold intensity needed to surpass the flux rate
between two SiC slabs held at a 300K temperature difference (light
blue curve) is roughly an order of magnitude larger, with $I \approx
10\mathrm{W}/\mu\mathrm{m}^2$. Finally, we remark that
\figref{plates}(g) also show the net flux rate achieved by replacing K
with Ag under the same lattice parameters and excitation conditions,
indicating signficant robustness with respect to the choice of
materials.

\section{Concluding Remarks}
\label{sec:conclusion}

We have presented a scheme for achieving large near-field thermal
energy transfer at the nanoscale based on the nonlinear
$\chi^{(3)}$-mediated interaction of thermal modes with externally
incident light. We employed a general coupled-mode framework to
predict power requirements and thermal flux rates in planar
configurations consisting of different (emitter and absorber)
materials separated by either a bulk or nanostructured nonlinear
medium, and in which resonantly enhanced incident light at
mid-infrared wavelengths upconverts mid-infrared thermal energy to
either near-infrared or visible wavelengths. We find that even when
the entire system is held at room temperature and at relatively low
pump intensities $\sim \mathrm{W}/\mu\mathrm{m}^2$, the rate of energy
upconversion can approach and even exceed
$10^{4}\mathrm{W}/\mathrm{m}^2$, which compares with typical flux
rates expected in the more common situation of identical slabs
separated by vacuum and held up to large $\sim 300$K temperature
differences. This scheme therefore not only provides a means to
achieve significant energy transfer at typically inaccessible
near-infrared or visible wavelengths, but also facilitates heat
exchange between dissimilar materials even when these are originally
in thermodynamic equilibrium. Finally, we remark that our specific
choices of planar geometries and materials are by no means optimal and
represent only a proof of concept. More importantly, while this scheme
could potentially be exploited in nanoscale radiative
cooling~\cite{shakouri2006nanoscale,chen2015heat,ding2016active} and
power generation~\cite{basu2009review,messina2012graphene}
applications, the efficiency and utility of the nonlinear process in
those situations will necessarily be degrared by heating introduced by
the pump and neglected in this work.  Such additional heating stems
from the presence of SPP resonances at the incident (mid-infrared)
wavelength, which necessarily cause conductive heating of the slab but
can potentially be mitigated by the introduction of a small vacuum gap
separating the nonlinear material from either or both of the
slabs. Similarly, instead of exploiting SPP resonant enhancement of
the incident light, one might also consider dielectric
resonances~\cite{quan2011deterministic,frank2010programmable} in
semi-transparent materials. A detailed analysis of the merits of this
scheme for radiative cooling or related applications that incorporates
pump-induced heating and related considerations will be the subject of
future work.

%Related ideas have been explored in the context of...

\section*{Funding}

This work was partially supported by the National Science Foundation
under Grant no. DMR-1454836 and by the Princeton Center for Complex
Materials, a MRSEC supported by NSF Grant DMR 1420541.

\section*{Acknowledgments}

We are thankful to Weiliang Jin and especially grateful to Riccardo
Messina for helpful comments and suggestions.

\end{document}